\documentclass[10pt]{emulateapj}
\usepackage{amsmath}
\usepackage{hyperref}
\usepackage{graphicx}
\usepackage{subfigure}
\usepackage{tabularx}

\hfuzz=\maxdimen
\newcommand{\pfrac}[2]{\left(\frac{#1}{#2}\right)}

\def\eps{\epsilon}

\def\fermi{{\em Fermi}}

\shorttitle{\fermi-LAT implications on IceCube neutrino origin}
\shortauthors{Wang, Zhao \& Li}

\begin{document}

\title{Implications of \fermi-LAT observations on the origin of IceCube neutrinos}

\author{Bin Wang\altaffilmark{1,2}, Xiaohong Zhao\altaffilmark{3,4} and Zhuo Li\altaffilmark{1,2,5}}

\altaffiltext{1}{Department of Astronomy, School of Physics, Peking
University, Beijing, China} \altaffiltext{2} {Kavli Institute for
Astronomy and Astrophysics, Peking University, Beijing, China}
\altaffiltext{3}{Yunnan Observatory, Chinese Academy of Sciences,
Kunming, China} \altaffiltext{4} {Key Laboratory for the Structure
and Evolution of Celestial Bodies, Chinese Academy of Sciences,
Kunming, China} \altaffiltext{5} {zhuo.li@pku.edu.cn}

\begin{abstract}
The IceCube (IC) collaboration recently reported the detection of
TeV-PeV extraterrestrial neutrinos whose origin is yet unknown. By
the photon-neutrino connection in $pp$ and $p\gamma$ interactions,
we use the \fermi-LAT observations to constrain the origin of the IC
detected neutrinos. We find that Galactic origins, i.e., the diffuse
Galactic neutrinos due to cosmic ray (CR) propagation in the Milky
Way, and the neutrinos from the Galactic point sources, may not
produce the IC neutrino flux, thus these neutrinos should be of
extragalactic origin. Moreover, the extragalactic gamma-ray bursts
(GRBs) may not account for the IC neutrino flux, the jets of active
galactic nuclei may not produce the IC neutrino spectrum, but the
starburst galaxies (SBGs) may be promising sources. As suggested by
the consistency between the IC detected neutrino flux and the
Waxman-Bahcall bound, GRBs in SBGs may be the sources of both the
ultrahigh energy, $\ga10^{19}$eV, CRs and the $1-100$~PeV CRs that
produce the IC detected TeV-PeV neutrinos.
\end{abstract}


\maketitle

\section{Introduction}
Recently, IceCube (IC) reports the detections of two PeV neutrinos
\citep{2PeV} and 26 sub-PeV additional events \citep{ic13} within
two years operation of IC-79 and IC-86.
In comparison with the expected number of 10.6 events from
atmospheric muons and neutrinos, the observed flux corresponds to an
excess with a signification of $4.3\sigma$ \citep{ic13}. This may
mark the first detection of high energy ($>$ TeV) extraterrestrial
neutrinos. Later on the three years of IC data improve the
signification up to $5.7\sigma$ \citep{IC14}. These neutrinos are
consistent with a flat energy spectrum, equal flavor ratio of 1:1:1
and isotropic sky distribution. The single flavor intensity of these
extraterrestrial neutrinos is
\begin{equation}
  E_\nu^2 \Phi_{\nu,\rm IC}\approx 10^{-8} \rm GeV cm^{-2}s^{-1}sr^{-1},
\end{equation}
corresponding to a $4\pi$ all sky integrated flux of
\begin{equation}
  E_\nu^2J_{\nu,\rm IC}\approx1.2 \times10^{-7}\rm GeVcm^{-2}s^{-1},
\end{equation}
from 60 TeV to 3 PeV energy range, and there is a lack of detected
$>$ 2 PeV neutrinos.

There are many scenarios that have been discussed regarding the
origin of these extraterrestrial neutrinos, both Galactic and
extragalactic models. The Galactic origins include the point source
contribution \citep{fox13}, and extended and diffuse sources due to
Galactic cosmic ray (CR) interaction with the interstellar medium
(ISM) during their propagation
\citep{gupta13,neronov13,ahlers13,razzaque13-1,joshi14,razzaque13-2,guo13}.
\cite{taylor14} even propose the diffuse neutrinos from an extended
Galactic halo. By assuming Galactic origin, the IC detected
neutrinos have been used to constrain the Galactic CR sources
\citep{a13}. On the other hand, extragalactic sources, e.g.,
gamma-ray bursts (GRBs) \citep{liuwang13,muraseioka13}, active
galactic nuclei (AGNs) \citep{stecker13,murase14}, and star forming
galaxies \citep{he13,liu13,tamborra14} have been discussed, as well
as extragalactic diffuse neutrinos due to CR propagation in cosmic
background photons \citep{laha13,roulet13,kalashev13}. The IC
detection, assuming extragalactic origin, has been used to constrain
the extragalactic CR source physics, e.g., the CR spectrum
\citep{murase13}, the production rate density \citep{katz13}, and
the physical condition of the CR accelerators \citep{winter13}.

The \fermi-Large Area Telescope (LAT) provides a survey of the
$\gamma$-ray sky from $30$ MeV to several hundred GeV with a
sensitivity more than an order of magnitudes surpassing its
predecessor EGRET. Many more point sources, as well as more precise
diffuse $\gamma$-ray background, have been detected by LAT. The
\fermi-Gamma-ray Burst Monitor (GBM) complements the LAT in its
observations of transient sources, especially gamma-ray bursts
(GRBs). In this paper we will use the $\gamma$-ray observations of
\fermi-LAT and GBM to constrain the Galactic and extragalactic
origins of the IC detected neutrinos, by assuming the $\gamma$-ray
and neutrino connection and extrapolation of the $\gamma$-ray
spectra.

The organization of the paper is as following. In section \ref{MW}
we discuss the Galactic models, including the diffuse neutrino
emission from CR interactions with ISM and extended halo matter
(section 2.1) and the neutrinos from Galactic point sources (section
2.2). Our constraint does not favor these Galactic sources. In
section \ref{XG}, we discuss the extragalactic model, especially the
GRB neutrino model. Combining with the LAT constraints of triggered
GRBs, we do not favor GRB model either (section 3.1). We further
propose that extragalactic neutrinos from AGN jets (section 3.2) and
SFGs (section 3.3) may be the possible source of IC neutrinos
(section 3.2). Finally section \ref{discussion} is conclusion and
discussion.

\section{Galactic origin} \label{MW}
The first question we need to ask about the IC detected neutrinos is
whether they can be produced in the Milky Way (MW), including the
contribution by point sources and the diffuse neutrinos from CR
propagation. Here we will derive the neutrino flux by extrapolation
of the $\gamma$-ray spectrum from \fermi-LAT observations, and then
compare it with the IC detected flux.

Both $\gamma$-rays and neutrinos can be produced by the interactions
between CR particles and medium matter ($pp$) or background photons
($p\gamma$). We can simply consider only $pp$ interactions and
neglect $p\gamma$ because the background photons are relatively rare
and $p\gamma$ time scale is much longer than $pp$ collisions.

In the case of $pp$ collisions, the flux ratio of $\pi^+$'s,
$\pi^-$'s and $\pi^0$'s is $\sim1:1:1$ at high energies. Neutrinos
are produced via charged pion's decay: $\pi^+\rightarrow
e^++\nu_e+\bar\nu_\mu+\nu_\mu$, $\pi^-\rightarrow
e^-+\bar\nu_e+\nu_\mu+\bar\nu_\mu$. Each neutrino carries one
quarter of the pion's energy . Photons are produced via neutral
pion's decay, $\pi^0\rightarrow\gamma+\gamma$, and each photon
carries one half of pion's energy. The flavor ratio of the produced
neutrinos is
$(\nu_e+\bar\nu_e):(\nu_\mu+\bar\nu_\mu):(\nu_\tau+\bar\nu_\tau)=1:2:0$,
and after oscillation the flavor ratio detected on earth becomes
$(\nu_e+\bar\nu_e):(\nu_\mu+\bar\nu_\mu):(\nu_\tau+\bar\nu_\tau)=1:1:1$
\citep{pdg12}. The number ratio of $\gamma$-rays and each flavor of
neutrinos generated via the processes above is $\sim 1:1:1:1$, so
the relation between the detected fluxes of diffuse $\pi^0$-decay
$\gamma$ rays and single flavor neutrinos at energies $E_\gamma
\simeq 2E_{\nu_\alpha}$ ($\alpha=e$, $\mu$ or $\tau$) is
\begin{equation}
E_\gamma^2\Phi_\gamma(E_\gamma)  \simeq
2E_{\nu_\alpha}^2\Phi_{\nu_\alpha}(E_{\nu_\alpha}).\label{eq2}
\end{equation}
Note that we neglect any attenuation of the $\gamma$-rays below 100
GeV, which we use, in the sources and during propagation.

\subsection{Cosmic ray propagation in the Milky Way}\label{MW:diffuse}

\begin{table}[t]
\caption{The $\gamma$-ray flux at 100GeV in different Galactic
regions observed by \fermi-LAT.}
\begin{center}
\begin{tabular}{ccc}
\hline
Galactic region & $E_\gamma^2\Phi_\gamma(\rm 100 GeV)$ & Solid Angle $\Delta\Omega$\\
 & $(\rm MeVcm^{-2}s^{-1}sr^{-1})$ &   (sr) \\
\hline
Local Galaxy & $3.552 \times 10^{-4}$ & $10.82$\\
Inner Galaxy & $3.748 \times 10^{-3}$ & $0.7773$\\
Outer Galaxy & $9.517 \times 10^{-4}$ & $0.9716$\\
\hline
\end{tabular}
\end{center}
\par
\tablecomments{The Galactic latitudes and longitudes of the three
defined regions are: $\left|b\right|>8^\circ$ (local Galaxy);
$\left|b\right|\leq8^\circ$ and $l < 80^\circ$ or $l > 280^\circ$
(inner Galaxy); $\left|b\right|\leq8^\circ$ and $80^\circ < l<
280^\circ$ (outter Galaxy).} \label{tab1}
\end{table}

\begin{figure}[bht]
\resizebox{\hsize}{!}{\includegraphics{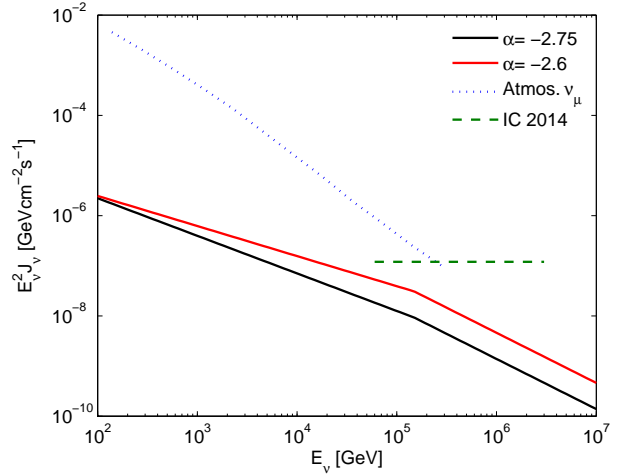}} \caption{The
extrapolated spectra of diffuse Galactic neutrinos. The red and
black lines represent the single flavor neutrino flux of
extrapolation with particle indices $-2.6$ and $-2.75$,
respectively. The green dashed line is the (all sky integrated) flux
of IC-detected extraterrestrial neutrinos, $E_\nu^2J_{\nu,\rm IC}$,
and the blue dotted line represents the background of atmospheric
muon neutrinos.} \label{spectrum}
\end{figure}

Consider the diffuse neutrino flux from the MW. After the Galactic
CRs escape from their sources, they propagate through the ISM. CR
particles are being scattered in the Galactic magnetic fields and
diffuse away from their sources, interacting with ambient gas. The
hadronic interactions produce not only neutrinos but also
$\gamma$-rays. Given the connection of the neutrino and $\gamma$-ray
flux, we can use the observed $\gamma$-ray flux to predict or
constrain the neutrino flux.

\fermi-LAT has provided a deep survey of the whole sky, from which
the diffuse Galactic emission (DGE) has been obtained by subtracting
the contribution from the detected point sources and the
instrumental and extragalactic background from the total flux
\citep{fermi}. the DGE consists of not only $\pi^0$-decay
$\gamma$-rays but also bremsstrahlung $\gamma$-rays of electrons and
positrons, IC $\gamma$-rays from electrons (positrons) scattering
cosmic microwave background (CMB) photons, and other components.
Here we consider the total DGE flux as the upper limit of
$\pi^0$-decay $\gamma$-ray flux.

Following \cite{fermi}, the entire sky can be dived into three
regions, so called "local Galaxy", "inner Galaxy", and "outer
Galaxy". The LAT $\gamma$-ray fluxes in different Galactic regions
are shown in Table \ref{tab1}, with the data taken from Figs 12, 15,
and 16 in \cite{fermi}, respectively. The all sky integrated energy
flux is the sum of the three regions, $\sum E_\gamma^2
{\Phi_\gamma}_i \Delta\Omega_i\simeq 7.47\times10^{-3} \rm
MeVcm^{-2}s^{-1}$ at 100GeV.

The observed CR spectrum \citep{CRreview} is roughly described as a
power law with two spectral breaks, i.e., the knee at $\sim3$PeV and
the ankle at $\sim5$EeV. The spectral index is $\sim -2.75$ below
the knee, and $\sim -3$ between the knee and the ankle. Beyond the
ankle the CR spectrum flattens with an index of $\sim-2.75$, and
then a cutoff appears at $\sim50$EeV. The ankle feature may suggest
an extragalactic CR component starts to dominates. Although the
exact CR energy where the transition from Galactic to extragalactic
CRs happen is under debates, it is generally believed that the CRs
below $\sim1$EeV are of Galactic origin.

Both the daughter $\gamma$-rays and neutrinos are roughly a constant
fraction of the primary protons, $E_\gamma\approx0.1E_p$ and
$E_\nu\approx0.05E_p$\footnote{Comes from the fact that each pion
carries $\sim$ 1/5 of the primary proton's energy.}. Thus the
$\gamma$-ray and neutrino spectra both follow that of the Galactic
CRs. We further assume that the CR spectrum anywhere in the Milky
Way is the same as the one observed on Earth. Therefore we set the
diffuse Galactic neutrino spectral index $\alpha = -2.75$ for
energies of 50GeV - 150TeV, and a fixed index $\beta=-3$ for
energies of 150TeV - 50PeV for simplification.

Assuming that the DGE all comes from $\pi^0$-decay photons, with
Equation \eqref{eq2} and the all sky integrated $\gamma$-ray flux
from Table \ref{tab1}, we obtain an upper limit to the flux of the
diffuse Galactic neutrinos, $E_\nu^2J_\nu=\sum E_\nu^2 {\Phi_\nu}_i
\Delta\Omega_i\simeq 3.74\times10^{-6}$$\rm GeVcm^{-2}s^{-1}$
($i=\rm Local,\,inner,\,ourter$) at $E_\nu=50$~GeV. The
extrapolation of neutrino flux from $E_\nu=50$~GeV to 50~PeV with
spectral profile assumed above is shown in Fig. \ref{spectrum},
which is significantly below the IC detected neutrino flux,
$E_\nu^2J_{\nu,\rm IC}$.

We also assume another harder CR spectrum with index of $-2.6$, then
the single-flavor diffuse neutrino flux at $E_\nu= 60$ TeV is
$E_\nu^2J_\nu=5.31 \times10^{-8}$$\rm GeVcm^{-2}s^{-1}$, which is
still about 3 times lower than the IC detected flux. Therefore the
IC detected neutrinos can not be produced by CR propagation in the
MW, unless the CR spectrum observed on Earth is not universal in the
MW, and can be much harder than $-2.7$, so that the DGE at
$E_\gamma>100$~GeV can be much harder as well. In order to account
for the IC detected neutrino flux at $E_\nu=1$~PeV, the $\gamma$-ray
spectrum should be extrapolated from 100 GeV with a photon index of
$\Gamma\sim-2.3$. Consider a spectral break, corresponding to the CR
knee, and the index of $\beta=-3$ above the knee, even harder
spectrum below the knee is required, $\Gamma\sim-2.2$.

\subsubsection{Galactic halo}
\cite{taylor14} recently propose that CRs produced by a
Galactic-center outflow may propagate into an extended Galactic halo
of a size $R_h\sim100$ kpc and a mass $M_h\sim10^{11}M_\odot$, and
lose most of their energy by $pp$ interactions. Given the IC flux
and the size, the total PeV-neutrino luminosity of the Galactic halo
will be $L_\nu\approx4\pi R_h^2E_\nu^2\Phi_\nu\approx10^{39}$erg
s$^{-1}$. According to our constraint, this requires that the DGE at
$\ga$TeV (without background radiation absorption) should be flatten
to be an index of $-2$. This is not in confliction with current
observations. However the following argument may not favor this
proposal for IC neutrinos.

Let us estimate the total neutrino flux from all the galaxies in the
universe since we expect the other galaxies, especially those
similar to the MW, also produce neutrinos in their halos. The
neutrino energy density in the universe can be estimated to be
$u_\nu\approx \xi_zL_\nu\rho_{\rm G}t_{\rm H}$, where $L_\nu$ is the
typical neutrino luminosity of each galaxy, $\rho_{\rm G}$ is the
galaxy number density, $t_{\rm H}$ is the Hubble time scale, and
$\xi_z$ accounts for the redshift evolution of the neutrino
production rate density in the universe. Thus the neutrino intensity
is $I_\nu=(c/4\pi)u_\nu=\xi_z(c/4\pi)L_\nu\rho_{\rm G}t_{\rm H}$.
The star formation rate (SFR) density in the local universe is
$\rho_{\rm SFR}=0.015M_\odot$yr$^{-1}$Mpc$^{-3}$ \citep{SFR}, while
the SFR in the MW is ${\rm SFR_{MW}}\approx2M_\odot$yr$^{-1}$
\citep{MWSFR}, thus we can estimate $\rho_{\rm G}\approx\rho_{\rm
SFR}/{\rm SFR_{MW}}\approx10^{-2}$Mpc$^{-3}$. If the neutrino
production rate evolves following the SFR density or the AGN
activity in the universe, then $\xi_z\sim3$ (\cite{wbbound}). Taking
$L_\nu\sim10^{39}$erg s$^{-1}$ and $t_{\rm H}\sim10$Gyr, we have
$I_\nu\sim3\times10^{-7}$GeV\,cm$^{-2}$s$^{-1}$sr$^{-1}$. This is
more than an order of magnitude larger than the IC observed flux.
Therefore in order for the IC excess being contributed by the
Galactic halo emitted neutrinos, our MW is required to be either an
unique galaxy or acting actively in a special phase.

\subsection{Galactic point sources}\label{subsection23}
Next consider the contribution of Galactic point sources by $pp$
interactions in the sources. \fermi-LAT has detected many new
sources in sky survey. The LAT 2-year Point Source Catalog (2FGL)
contains 1873 high energy $\gamma$-ray sources detected by LAT
during the period of August 4, 2008 to July 31, 2010
\citep{nolan12}. Among these sources, there are 195 Galactic
sources, 1102 extragalactic sources, and 576 unknown sources (US).
The spectral shapes of these sources are divided into three types:
power law, pow law with exponential cutoff, and log-parabola. We
calculate photon fluxes from each source at 100GeV with the given
spectral parameters in 2FGL. The total fluxes for different types of
sources at 100GeV are shown in Table \ref{tab2}. We obtain that the
total fluxes of identified Galactic and US at $E_\gamma=100$~GeV are
$E_\gamma^2J_{\gamma,\rm MW}=1.29\times10^{-7}\rm
GeV\,cm^{-2}s^{-1}$ and $E_\gamma^2J_{\gamma,\rm
MW}=2.58\times10^{-7}\rm GeV\,cm^{-2}s^{-1}$, respectively.

\begin{table}[t]
\caption{$\gamma$-ray fluxes of 2FGL sources at 100GeV.}
\begin{center}
\begin{tabular}{ccc}
\hline
Source type & $E_\gamma^2J_\gamma$(100GeV)$/\rm GeVcm^{-2}s^{-1}$\\
\hline
Galactic & $1.29\times10^{-7}$\\
spp & $5.77\times10^{-8}$\\
pwn & $4.04\times10^{-8}$\\
psr & $1.39\times10^{-8}$\\
snr & $1.21\times10^{-8}$\\
glc & $4.01\times10^{-9}$\\
hmb & $5.06\times10^{-10}$\\
nov & $1.60\times10^{-13}$\\
US & $2.58\times10^{-7}$\\
extragalactic & $1.43\times10^{-6}$\\
\hline
\end{tabular}
\end{center}
\par
\tablecomments{Following the 2FGL, the abbreviations used for the
Galactic source types are: spp: special case (potential association
with SNR or PWN); pwn: pulsar wind nebula; psr: pulsar (include both
those identified by pulsations and those no pulsations seen in LAT
yet); snr: supernova remnant; glc: globular cluster; hmb: high-mass
binary; and nov: nova.} \label{tab2}
\end{table}

A significant fraction of the USs may be Galactic other than
extragalactic sources. We estimate the contribution of those USs
that are of Galactic origin to the $\gamma$-ray flux on Earth.

\begin{figure*}[bpt]
  \begin{center}
    \includegraphics[width=3.3in]{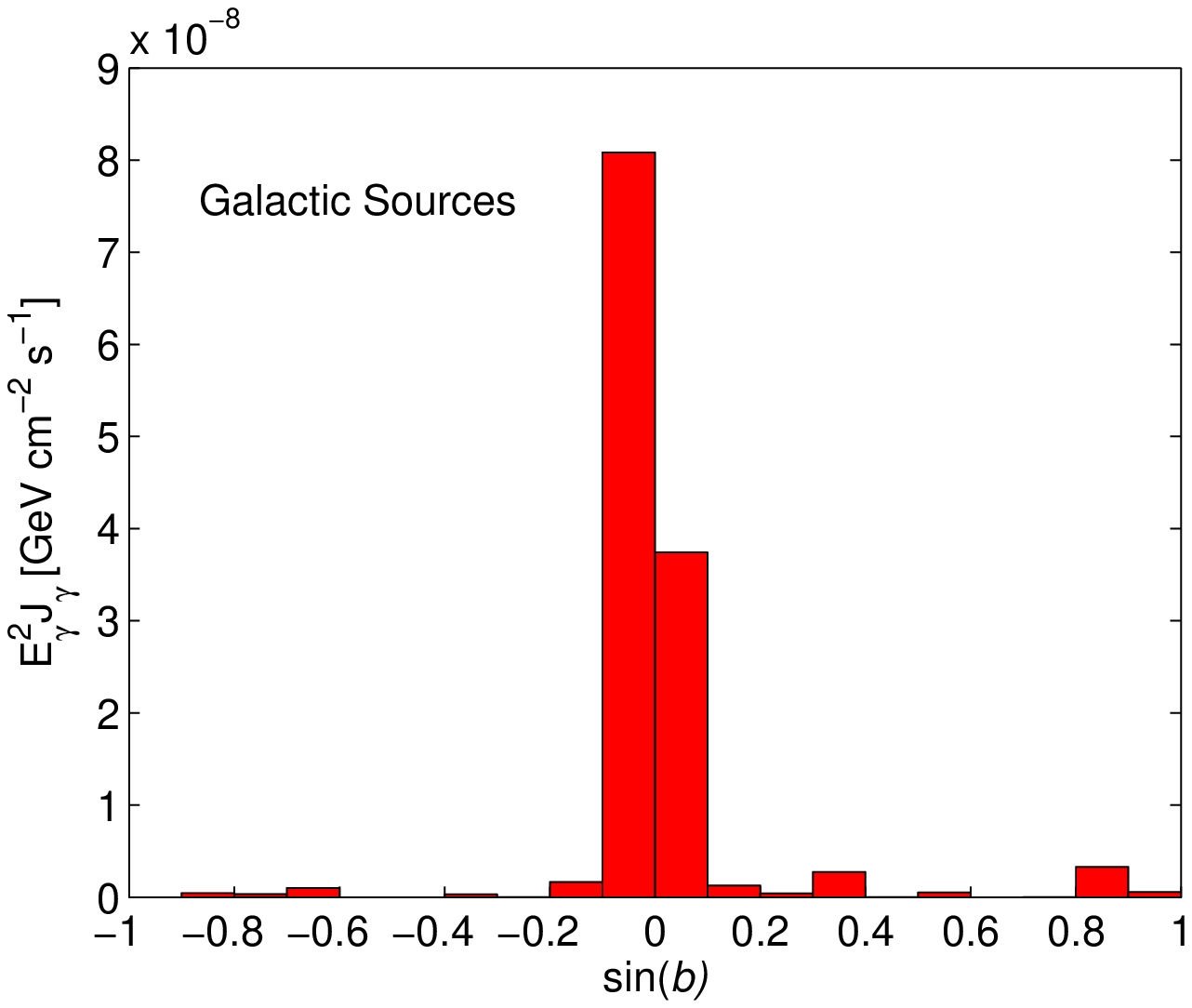}
    \includegraphics[width=3.3in]{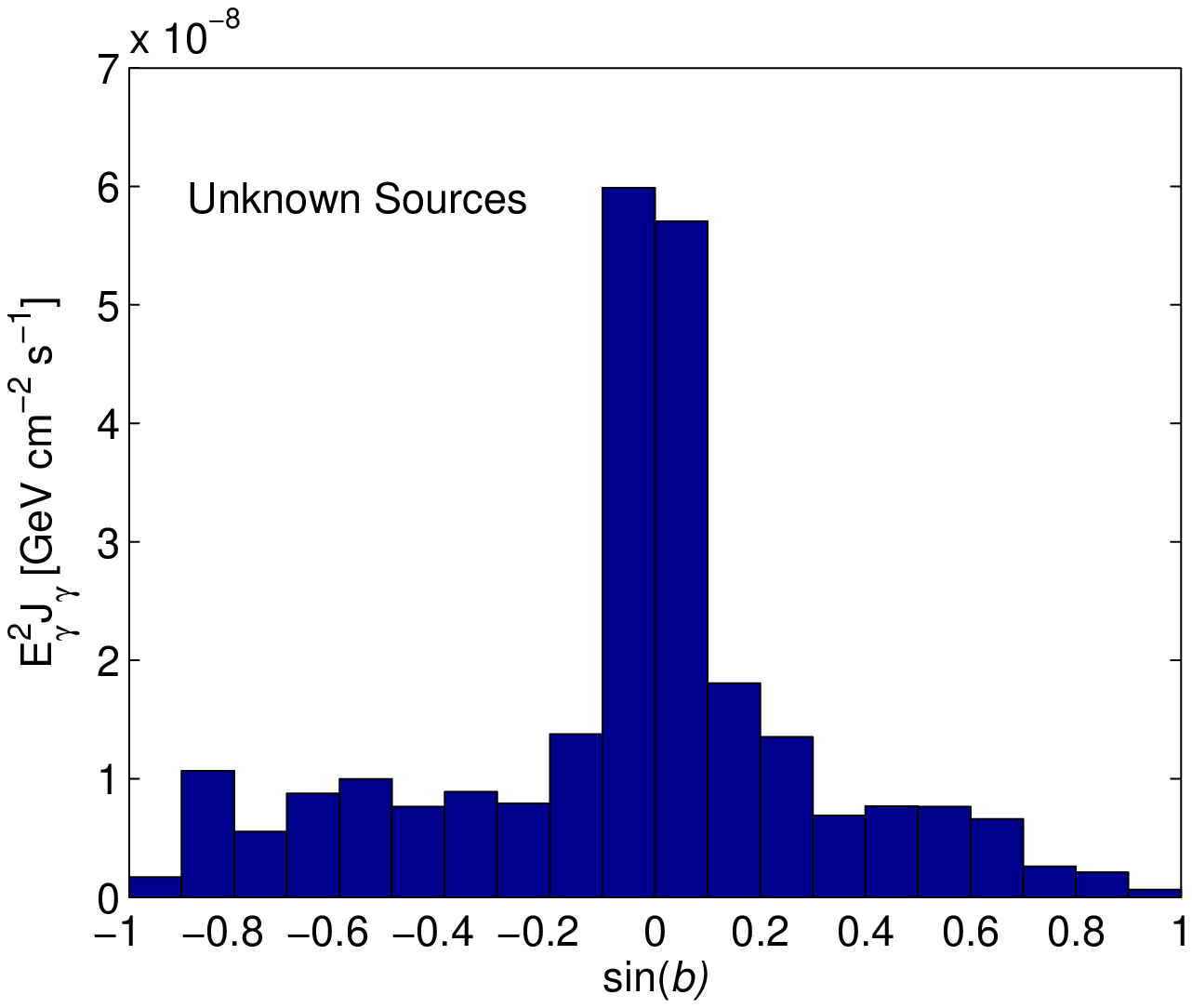}\\
  \end{center}
\caption{The Galactic latitude distribution of the $\gamma$-ray flux
at 100 GeV from identified Galactic sources ({\em left}) and USs
({\em right}). The point sources are separated into bins with equal
$\Delta\sin(b)$, so that the isotropic distribution corresponds to
an uniform distribution along $\sin(b)$ bins.}
  \label{fig2}
\end{figure*}

According to 2FGL, we show the Galactic latitude distribution of
$\gamma$-ray flux of identified Galactic sources and USs at
$E_\gamma=100$~GeV in Fig. \ref{fig2}. The US distribution consists
of Galactic and extragalactic components. We assume that the
Galactic latitude distribution of the Galactic USs follows the same
shape of the identified Galactic sources, and that the extragalactic
one is isotropically distributed. We should subtract the isotropic
extragalactic component from the total US flux to obtain the flux of
Galactic USs. By comparing the latitude distributions of the
identified Galactic sources and the USs, we can find that the
emission at $|\sin b|>0.1$ is dominated by extragalactic component.
Subtracting an isotropic background flux to all the $\sin b$ bins,
the expected contribution of USs to the Galactic point source flux
is $E_\gamma^2J_\gamma\sim 1\times10^{-7}\rm GeVcm^{-2}s^{-1}$ at
$E_\gamma=100$~GeV. So the total $\gamma$-ray flux of all Galactic
point sources (including identified sources and USs) is
$E_\gamma^2J_\gamma\approx 2.3\times10^{-7}\rm GeVcm^{-2}s^{-1}$.
With Eq. \eqref{eq2}, the neutrino flux at $E_\nu=50$~GeV is
$E_\nu^2J_\nu\sim 1.1\times10^{-7}\rm GeVcm^{-2}s^{-1}$.

We assume that the neutrino spectrum is flat $E_\nu^2\Phi_\nu\propto
E_\nu^0$. This may be an optimistic estimate of the neutrino flux
because CR spectrum is expected to be flat in a strong shock, and
the secondary $\gamma$-ray and neutrino spectra follow the CR
spectrum. So the expected PeV neutrino flux from Galactic point
sources is
\begin{equation}\label{eq:gps}
   E_\nu^2J_{\nu,\rm point}\approx1.1\times10^{-7}\rm GeV cm^{-2}s^{-1},
\end{equation}
not larger than the IC excess.

Following the discussion by \cite{nolan12} (their section 5.4), the
contribution of USs to the Galactic source neutrino flux can be even
lower, and the above conclusion may be more robust: (1) Despite the
flux of USs increases sharply in Galactic plane, it is attributable
to the relative lack of sources at $\left|b\right|<10^\circ$ in many
of the extragalactic source catalogs used for source identification;
(2) The USs in galactic plane with curved spectra tend to cluster in
bright Galactic diffuse emission regions, suggesting that at least a
fraction of them may be DGE maxima, that are not adequately modeled
by DGE model; (3) Some USs may be unreal because there are much more
fraction (51\%) of USs with doubt being a source in contrast to the
fraction of identified sources (14\%). Moreover, the $\gamma$-ray
flux may be dominated by electron emission, other than $pp$
interactions. In conclusion, the flux of PeV neutrinos from Galactic
point sources may be far below that in equation (\ref{eq:gps}), and
may not reach the IC excess flux.

\section{Extragalactic origin}
\label{XG}

Since Galactic origins of IC neutrinos are not favored, we turn to
discuss the extragalactic origin.

The scenario that the IC neutrinos are cosmogenic neutrinos produced
via $p\gamma$ interactions between CR particles and cosmic
background photons is disfavored, because normalizing the expected
neutrino flux to that observed by IC at 1~PeV leads to an over
predicted neutrino flux at EeV range
\citep{roulet13,laha13}\footnote{CR propagation in CMB may still
produce IC neutrinos in some specific cases \citep{kalashev13}.}. We
discuss the other extragalactic neutrinos sources below, i.e., GRBs,
AGN jets, and starburst galaxies (SBGs).

\subsection{Gamma-ray bursts}
GRBs have long been proposed to be a strong candidate of the source
of ultrahigh energy (UHE) CRs \citep{waxman95}, and the $p\gamma$
collisions are expected to produce intense neutrino emission around
PeV range \citep{waxman97}, consistent with the current IC
constraint on the neutrino spectrum. It is interesting to check
whether GRB neutrinos can account for the IC diffuse neutrinos.
In the following, assuming first that GRBs are the IC neutrino
source, we estimate the average neutrino flux of a triggered GRB,
which should be compared to the upper limit IC puts on the triggered
GRBs. By doing this, we should assume that {\em the neutrino flux
from a GRB is proportional to the MeV-range flux.}

For a $\gamma$-ray detector monitoring the whole sky with a
sensitivity of $p_{\rm th}$, the GRB trigger rate is
\begin{equation}
  \dot{N}_{\rm trig}=\int_0^{z_{\max}}\frac{R(z)}{1+z}\frac{dV}{dz}dz\int_{p_{\rm th}4\pi
  D_L^2k(z)}^\infty\phi(L)dL,
\end{equation}
where $R(z)$ is the redshift dependence of GRB rate density,
$\phi(L)$ is the GRB luminosity function, $D_L$ is the GRB
luminosity distance, $dV/dz$ is the volume-redshift relation of the
universe, and $k(z)$ corresponds to $k$-correction, depending on GRB
spectrum and detector energy range (see below). The (time averaged)
$\gamma$-ray flux from triggered GRBs is
\begin{equation}
  \Phi_{\rm trig}=\int_0^{z_{\max}}\frac{R(z)}{4\pi D_L^2(1+z)}\frac{dV}{dz}dz\int_{p_{\rm th}4\pi
  D_L^2k(z)}^\infty E(L)\phi(L)dL,
\end{equation}
whereas the total one, including the contribution from untriggered
GRBs is
\begin{equation}
  \Phi_{\rm tot}=\int_0^{z_{\max}}\frac{R(z)}{4\pi D_L^2(1+z)}\frac{dV}{dz}dz\int_0^\infty
  E(L)\phi(L)dL.
\end{equation}
Here $E$ is the GRB energy, for which we simply assume $E\propto L$.

Recently the GRB rate density and luminosity function have been well
constrained by using the large sample of {\em Swift} GRBs with
redshift measurement \citep{GRBrate}:
\begin{equation}
  R(z)=R(0)\left\{\begin{array}{ll}
  (1+z)^{n_1} & z<z_1\\
  (1+z_1)^{n_1-n_2}(1+z)^{n_2} & z\ge z_1\end{array}\right.
\end{equation}
\begin{equation}
  \phi(L)=\phi_0\left\{\begin{array}{ll}
  (L/L_*)^x & L<L_* \\
  (L/L_*)^y & L\ge L_* \end{array}\right.
\end{equation}
where $R(0)=0.84 \rm Gpc^{-3}yr^{-1}$, $z_1=3.6$, $n_1=2.07$,
$n_2=-0.7$, $L_*=10^{52.05} \rm erg\,s^{-1}$, $x=-0.65$, $y=-3$, and
$\phi_0$ is defined such that $\int_0^\infty\phi(L)dL\equiv1$. For
\fermi-GBM the threshold is $p_{\rm th}=0.71 \,\rm photons~
cm^{-2}s^{-1}$ in the energy range of 50-300 keV \citep{meegan09}.
In this case we write
\begin{equation}
  k(z)=\frac{\int_{1keV}^{10MeV}\eps
  n(\eps)d\eps}{\int_{50keV(1+z)}^{300keV(1+z)}n(\eps)d\eps},
\end{equation}
where $n(\eps)$ is the GRB spectrum in the rest frame, for which we
assume a broken power law with photon indices of $\alpha=-1$ and
$\beta=-2.2$ and a sharp break at $\eps_{\rm break}=511$ keV. We
will take $z_{\max}=8$ and assume a flat $\Lambda$CDM universe with
$\Omega_m=0.27$, $\Omega_\Lambda=0.73$, and $H_0=71 \rm
km\,s^{-1}Mpc^{-1}$.

The GBM trigger threshold is $p_{\rm th}=$ 0.71 photons $\rm
cm^{-2}s^{-1}$ \citep{meegan09}. There were 183 GRBs triggered
between 2008 July 11 and 2009 March 31, corresponding to a GRB
trigger rate of $\sim 260$ burst $\rm yr^{-1}$ \citep{meegan09}.
Taking into account the Earth occultation and the South Atlantic
Anomaly passage, only 65\% of GRBs above the GBM threshold can be
detected \citep{liuwang13}. So for a detector monitoring all sky
with the GBM threshold, the trigger rate should be $\dot{N}_{\rm
trig}\approx400yr^{-1}$. On the other hand, if $p_{\rm th}=$0.7,
0.5, and 0.3 photons $\rm cm^{-2}s^{-1}$, we can calculate with the
above formula that $\dot{N}_{\rm trig}=$220, 306, and 480 $yr^{-1}$,
and $\Phi_{\rm trig}/\Phi_{\rm tot}=$0.71, 0.75, and 0.8,
respectively. To obtain the GBM trigger rate $\dot{N}_{\rm
trig}=400yr^{-1}$ one needs $p_{\rm th}=0.37$ photons $\rm
cm^{-2}s^{-1}$.

Taking $\dot{N}_{\rm trig}=400$ yr$^{-1}$ and $\Phi_{\rm
trig}/\Phi_{\rm tot}=0.7$, if GRBs can account for the IC neutrino
flux $f_{\rm IC}=E_\nu^2J_{\nu,\rm IC}\rm
\log(2PeV/60TeV)=2.3\times10^{-7}$GeV\,cm$^{-2}$s$^{-1}$ ($4\pi$
integrated and single flavor), the average neutrino fluence of a
GBM-triggered GRB is required to be
\begin{equation}
F_{\rm trig}=\frac{f_{\rm IC}}{\dot{N}_{\rm trig}}\frac{\Phi_{\rm
trig}}{\Phi_{\rm tot}}=1.3\times 10^{-2}\frac{\Phi_{\rm
trig}}{0.7\Phi_{\rm tot}}\frac{400{\rm yr}^{-1}}{\dot{N}_{\rm trig}}
\rm GeVcm^{-2}.\label{GRBnuflux}
\end{equation}
The $p\gamma$ interactions produce not only neutrinos but also
$\gamma$-rays, which generate electromagnetic cascade emission in
GeV energy range, which can be observed by \fermi-LAT. Using the
neutrino and $\gamma$-ray connection, the \fermi-LAT observations of
GRBs help to constrain that the average neutrino fluence from a GBM
triggered GRB is \citep{li13}
\begin{equation}
  F_{\rm LAT-bound}\sim\rm 2\times 10^{-3}GeVcm^{-2},
\end{equation}
smaller than the required neutrino flux (equation \ref{GRBnuflux}).
The IC has also given an upper limit to the neutrino fluence from a
triggered GRB (averaged over 215 GRBs), $F_{\rm IC-bound}\approx
E^2F_\nu\times\log(10)/215=7\times10^{-4}\rm GeVcm^{-2}$ (using
$E^2F_\nu\approx0.15\rm GeVcm^{-2}$; see Fig.1 in \cite{ic12}). This
is also smaller than required, although the comparison is not
straightforward because these 215 GRBs include not only those
detected by GBM but also the other detectors.

Thus we reach the conclusion that GRB neutrinos may not account for
the IC detected neutrinos (though the other GRB neutrino models that
cannot be constrained by the neutrino-$\gamma$-ray connection may
still work \citep{muraseioka13}).

It may be noted that the reported GBM threshold and GRB detection
rate seem not completely consistent with the GRB redshift and
luminosity distributions derived by \cite{GRBrate}. However our
result of using $\dot{N}_{\rm trig}=400$ yr$^{-1}$ and $\Phi_{\rm
trig}/\Phi_{\rm tot}=0.7$ is robust since $\Phi_{\rm trig}/\Phi_{\rm
tot}\sim0.7$ is not sensitive to $p_{\rm th}$ and taking
$\dot{N}_{\rm trig}\sim200$ yr$^{-1}$ (for GBM threshold value) even
enhances the neutrino emission (eq \eqref{GRBnuflux}).

Our conclusion is similar to \cite{liuwang13}, but the main
difference in between is the following. \cite{liuwang13} use several
assumptions of the GRB model, e.g., the jet Lorentz factor, the
variability timescale, the fraction of energy in accelerated
protons, etc., in order to calculate the neutrino production. Here
we only need to assume that the neutrino flux is proportional to the
$\gamma$-ray one (eq \eqref{GRBnuflux}). This may be reasonable
given that the MeV $\gamma$-rays essentially carry away all the
energy of electrons which probably carry some constant fraction of
the total jet energy, and that the neutrinos carry away some
constant fraction of the energy of protons which also may carry some
constant fraction of the total jet energy. These may be true in a
statistical point of view, although it may not hold for individual
GRBs.

\subsection{Jets of active galactic nuclei}
AGN jets have long been predicted to be high energy CR and neutrino
sources, and the dominant contribution of neutrino emission may be
quasar hosted blazars, in particular, the flat spectrum radio
quasars \citep[FSRQ; e.g.,][]{murase14}, where the high energy
neutrino production is due to photopion interactions between jet
produced CRs and the external broadline and dust radiation.

FSRQs are also bright in $\gamma$-ray emission, which is possible to
be produced by the primary electrons accelerated in the jets
accompanying the production of high energy CRs. We may use the
\fermi-LAT observations of FSRQs to make constraint on neutrino
production. \cite{fsrq} has reported the luminosity function and
redshift evolution of the \fermi detected FSRQs, which imply that
the diffuse $\gamma$-ray flux from FSRQs is $4.1\times10^{-6}\rm
MeV\,cm^{-2}s^{-1}sr^{-1}$ from $\sim0.1$~MeV to $10$~GeV range
(Figure 11 therein), i.e., a whole sky integrated flux of
\begin{equation}
  f_{\gamma}=5.1\times10^{-5}\rm GeV\,cm^{-2}s^{-1}.
\end{equation}
Compared with IC flux we have the ratio of neutrino to $\gamma$-ray
flux, $r_{\nu/\gamma}=3f_{IC}/f_\gamma\simeq1.4\times10^{-2}$, where
the factor 3 comes from the equal flux in the three neutrino
flavors. This ratio is consistent with the estimated photomeson
production efficiency for CR protons above the threshold for
interacting with the broadline emission,
$f_{p\gamma}\sim5\times10^{-2}f_{\rm cov,-1}L_{\rm AD,46.5}^{1/2}$,
where $f_{\rm cov}$ is the cover factor of the broadline emission
region and $L_{\rm AD}$ is the accretion disk luminosity
\citep{murase14}, if jet-produced high energy electrons and CRs have
comparable energies and a significant fraction of CRs lies above the
threshold of photopion production. Therefore, FSRQs may produce
diffuse neutrinos with a flux comparable to IC detection.

However, AGN jets may have difficulty in producing the detected,
flat neutrino spectrum from tens TeV to few PeV
\citep{winter13,murase14}. Because the low energy radiation peaks at
infrared to UV range, the photopion interactions tend to produce
high neutrino flux at $>$PeV, in contrast to the IC observation
which appears as lack of neutrinos above few PeV. Because of the
decreasing radiation above UV frequency, the predicted neutrino flux
decreases fast below PeV range, also in contrast to the IC
observation. Thus the $\la100$TeV neutrinos may need the other
origins instead of AGNs, and future observations at EeV range are
needed to test the high neutrino flux from AGN jets.

\subsection{Starburst galaxies}
Starburst galaxies (SBGs) have been expected to be promising
neutrino sources \citep{loebwaxman}, given that they are strong CR
sources and that the high density ISM and high magnetic field lead
to high efficiency of CR energy loss by $pp$ collisions. CRs at
$\la100$~PeV may significantly lose their energy \citep{loebwaxman}.
It is noticed that the IC neutrino flux is well consistent with the
Waxman-Bahcall bound \citep{wbbound}, which may suggest that all the
CR energy is lost in pion production.

\fermi-LAT has detected several SBGs in $0.1-100$~GeV
range\citep{fermi-sfg}. By comparing with their SFRs estimated by
radio and far infrared detections, we have the $\gamma$ ray
luminosity and SFR relation in SBGs \citep{fermi-sfg,katz13}
\begin{equation}
  \nu
L_\nu({\rm GeV})/{\rm SFR}\approx10^{46}{\rm erg}/M_\odot
\end{equation}
(where $L_{0.1-100\rm GeV}\sim7\nu L_\nu(\rm GeV)$ is used).
Assuming this relation to be universal, the GeV $\gamma$-ray
production rate density in the local universe is
$E_\gamma^2Q_{\gamma}({\rm GeV})=\rho_{\rm SFR}(\nu L_\nu({\rm
GeV})/{\rm SFR})\approx1.5\times10^{44}$erg\,yr$^{-1}$Mpc$^{-3}$.
The GeV $\gamma$-ray intensity (without attenuation) is
$E_\gamma^2\Phi_\gamma({\rm GeV})=\xi_zt_{\rm
H}(c/4\pi)E_\gamma^2Q_\gamma({\rm
GeV})\approx3.4\times10^{-7}(\xi_z/3)$GeV\,cm$^{-2}$s$^{-1}$sr$^{-1}$.
The neutrino and $\gamma$-ray connection in $pp$ collisions leads to
the neutrino intensity in GeV range of $E_\nu^2\Phi_\nu({\rm 0.5
GeV})=(1/2)E_\gamma^2\Phi_\gamma({\rm
GeV})=1.7\times10^{-7}(\xi_z/3)$GeV\,cm$^{-2}$s$^{-1}$sr$^{-1}$,
which is one order of magnitude higher than the IC detected flux at
PeV scale. However, as suggested by CR confinement time ($\propto
E_p^{-0.5}$) and CR spectral slope ($\propto E_p^{-2.7}$)
measurement of CRs in MW, the spectrum of injected CRs may be
$dn_p/dE_p\propto E_p^{-2.2}$. If the CRs lost most of their energy
in $pp$ interactions then the neutrino spectral slope follows that
of the CRs, and the neutrino flux at PeV scale extrapolated from GeV
range is
\begin{equation}
  E_\nu^2\Phi_\nu\approx10^{-8}\frac{\xi_z}{3}\pfrac{E_\nu}{{\rm 1 PeV}}^{-0.2}\rm
  GeV\,cm^{-2}s^{-1}sr^{-1}.
\end{equation}
consistent with IC detection. Moreover, two SBGs, NGC253 and M82,
have been detected in TeV range, which show TeV flux lower than GeV
one by about one order of magnitude \citep{fermi-sfg}. Thus
\fermi-LAT observations suggest that neutrinos from $pp$ interaction
in SBGs may account for the IC detection.

The calculation above does not consider that the local SFR density
is dominated by normal star forming galaxies other than SBGs.
However it is suggested in observations that most of the stars in
the universe formed in SBGs at high redshift $z\sim2-4$
\citep{SF_sbg1,SF_sbg2}, thus the neutrino production is dominated
by SBGs at $z\ga2$. The above calculation is available since the
local neutrino production does not contribute significantly to the
total neutrino flux.

\section{Conclusion and Discussion} \label{discussion}

Using \fermi-LAT observations and with the neutrino and $\gamma$-ray
connection, we have constrained the origin of IC neutrinos. The main
conclusions are below:

First, the diffuse Galactic neutrino emission from CR propagation in
MW cannot account for the IC detected neutrino flux if the CR
spectral slope $\propto E_p^{-2.75}$ is universal in MW. In order to
account for the IC neutrinos at PeV scale, the DGE spectral slope at
$>100$~GeV should be harder than $\Gamma\sim-2.3$.

We obtain that the upper limit to the diffuse Galactic neutrino flux
at 60 TeV is $\sim3$ times lower than the IC excess by considering
that the total \fermi-LAT detected emission is from $pp$
interactions. However, $pp$ interactions only contribute a fraction
of the DGE. By the modeling of \cite{fermi}, the $\pi^0$-decay
photons contribute to $\sim1/3$ of the total LAT  flux in the
``local" and ``outer Galaxy", and $\sim1/2$ of the total LAT flux in
the  ``inner Galaxy". The expected diffuse Galactic neutrino flux
should be at least a factor of $\sim2$ lower than the upper limit we
obtain, so the diffuse Galactic neutrino flux may contribute to
$\sim10\%$ of the IC detected neutrino flux.

A study of the Galactic latitude distribution of the detected
neutrinos would be more powerful test than only considering the
total neutrino flux, but needs much more statistics of neutrino
events. Future IC detection of the latitude distribution should be
compared with the prediction of \cite{stecker79}. On the other hand,
the detection of diffuse PeV photons would be more direct test
\citep{ahlers13}, other than extrapolation of $\gamma$-ray spectrum
from GeV to PeV scale. However, the current TeV-PeV photon
detections only cover limited parts of the sky, e.g., biased in the
Northern Hemisphere, in contrast to \fermi-LAT's deep survey of the
whole sky.

Second, the high energy $\gamma$-ray point sources in MW cannot
account for the IC excess, unless the $\gamma$-ray spectra of these
sources at $>100$~GeV is unexpectedly harder than a flat spectrum
with photon index $\Gamma=-2$.

The point source spectral indices beyond 100 GeV are the main
uncertainty. However, photons from some types of sources, such as
pulsars, are not hadronic dominant at 100GeV, which further reduces
the expected flux from point sources. Moreover, the Galactic point
sources concentrate on the Galactic disk, very different from the
sky location of IC neutrinos, which is consistent with isotropic
distribution. It should also be mentioned that although the neutrino
flux from Galactic sources can hardly reach IC excess, both are in
the same order of magnitude, suggesting that the Galactic sources
may contribute to a fraction of IC neutrinos.

Again, the Galactic latitude distribution of detected neutrinos is
more powerful and straightforward test to the Galactic point source
origin.  The IC-detected neutrinos arriving from high Galactic
latitudes seem to disfavor the Galactic point source origin, but
more detections of neutrinos in the future are required to study the
latitude distribution with high confidence level.

Third, neutrino productions in GRB jets may not account for the IC
neutrino flux. This is based on the assumption that in GRBs the
neutrino flux is proportional to the $\gamma$-ray flux. We have used
the LAT observations of GRB GeV emission to constrain neutrino flux.
As time goes by, IC collects more observational results on GRBs, the
constraint on GRB neutrino will be more and more stringent.

Our method is applied to the classic GRBs with the common picture
that the neutrino production occurs in the region where the main
burst of MeV $\gamma$-rays are produced, e.g., the internal shock
region \citep{waxman97}. Thus we do not constrain the neutrino
production when the jet is still deep inside the GRB progenitor
\citep{mw01}. \cite{muraseioka13} find that low-power jets inside
progenitors of GRBs may produce higher flux of TeV-PeV neutrinos. It
would be important to measure the emissivity in the universe by more
observations of these ``low power GRBs".

Forth, \fermi-LAT observation suggests that AGN jets may produce
neutrino flux as high as the IC flux. However, AGN jets may have
difficulty in explaining the flat spectrum from tens TeV to few PeV
detected by IC. AGN jets may not account for the tens-TeV neutrinos
detected by IC, and the future EeV neutrino experiments would be
important to test the predicted AGN neutrino flux.

For the AGN jet produced neutrinos to reach the IC detected flux,
their local universe CR generation rate should be $10-100$ times
larger than the local UHE CR emissivity, because of the low
photopion production efficiency \citep{murase14,dermer14}. This is
in contradiction with the consistency between IC detected neutrino
flux and Waxman-Bahcall bound, unless that CR energy production rate
decreases sharply by $10-100$ times from $\sim100$ PeV to
$\sim10^{19}$eV.

Finally, we use the \fermi-LAT detections of individual SBGs to
constrain the PeV neutrino flux from SBGs, and find that they can
account for the IC excess. \cite{liu13} have considered the neutrino
emission from star forming galaxies, including SBGs, to explain the
IC neutrinos. They use a more specific model, instead of the
$\gamma$-ray and neutrino connection as we emphasize here.

It should be noted that \cite{wbbound} have derived an maximum
diffuse neutrino flux by assuming all CRs lose energy in pion
productions and normalizing the neutrino flux to the UHE CR
production rate density. The 60TeV-2PeV neutrino flux detected by IC
turns out to match the predicted Waxman-Bahcall bound, which implies
that the CRs in $1-100$~PeV is the same component as the UHE CRs
\citep{katz13} and all the CR energy is lost in pion
production.\footnote{\cite{liu13} do not need CRs losing all their
energy inside the galaxies, because they do not normalize the CR
flux in $1-100$~PeV range to the observed UHE CR one, i.e., they are
different CR components of different origins in the universe. The
consistency of the IC neutrino flux with the Waxman-Bahcall bound
happens to be so.}

As there is no bright AGNs in the local universe within the UHE CR
energy loss length ($\sim100$ Mpc; due to interaction with cosmic
microwave background photons), GRBs are the more promising sources
for UHE CRs. Thus, a likely explanation to the IC neutrinos is that
the CRs corresponding to the IC neutrinos are also produced by GRBs.
These CRs do not lose significant fraction of their energy in GRB
jets, as constrained by \fermi-LAT and IC observations of GRBs, but
they lose most energy after escaping from GRB jets and propagate in
GRB host galaxies, which are mostly SBGs. Future deeper observations
of high energy $\gamma$-rays and neutrinos from individual GRBs and
SBGs by, e.g., CTA and IC, etc, can test this interpretation
\cite[e.g.,][]{ic-point}.

\begin{acknowledgments}
The authors thank the referee for helpful comments, and A. W. Strong
and G. J\'{o}hannesson for helpful discussions. This work is
supported by NSFC (11273005, 11203067), SRFDP (20120001110064), the
973 Program (2014CB845800), Yunnan Natural Science Foundation
(2011FB115), and the West Light Foundation of CAS.
\end{acknowledgments}

\end{document}